# Microresonator devices lithographically introduced at the optical fiber surface

N. Toropov,[1†*] S. Zaki,[1] T. Vartanyan,[2] and M. Sumetsky[1]

[1]*Aston Institute of Photonic Technologies, Aston University, Birmingham B4 7ET, UK*
[2]*ITMO University, St. Petersburg 197101, Russia*
[†] *Current address: Physics and Astronomy, University of Exeter, Exeter EX4 4QD, United Kingdom*
*\*Corresponding author: n.toropov@exeter.ac.uk*

**We present a simple lithographic method for fabrication of microresonator devices at the optical fiber surface. First, we undress the predetermined surface areas of a fiber segment from the polymer coating with a focused $CO_2$ laser beam. Next, using the remaining coating as a mask, we etch the fiber in a hydrofluoric acid solution. Finally, we completely undress the fiber segment from coating to create a chain of silica bottle microresonators with nanoscale radius variation (SNAP microresonators). We demonstrate the developed method by fabrication of a chain of five 1 mm long and 30 nm high microresonators at the surface of a 125 micron diameter optical fiber and a single 0.5 mm long and 291 nm high microresonator at the surface of a 38 micron diameter fiber. As another application, we fabricate a rectangular 5 mm long SNAP microresonator at the surface of a 38 micron diameter fiber and investigate its performance as a miniature delay line. The propagation of a 100 ps pulse with 1 ns delay, 0.035$c$ velocity, and negligible dispersion is demonstrated. In contrast to the previously developed approaches in SNAP technology, the developed method allows the introduction of much larger fiber radius variation ranging from nanoscale to microscale.**

Microresonators supporting whispering gallery modes (WGMs) are versatile photonic devices explored in fundamental and applied research ranging from cavity quantum electrodynamics and optomechanics to microwave photonics and biomedical sensing, finding applications as miniature optical delay lines and buffers, optical switches, label-free sensors, microlasers, nonlinear parametric converters, and optical frequency comb generators [1–9]. Microresonator-based photonic circuits with various levels of functionality have been demonstrated using different lithographic fabrication platforms achieving excellent fabrication precision reaching a few nanometers and low insertion loss [10-13]. However, further development of several important applications of resonant photonic devices, such as miniature delay lines and frequency comb generators, requires orders of magnitude better (i.e., subangstrom) fabrication precision. Such outstanding precision is enabled by the SNAP (Surface Nanoscale Axial Photonics) platform [14-17]. In the SNAP technology, optical WGM microresonator structures are formed at the surface of an optical fiber by its nanoscale deformation. The breakthrough performance of resonant SNAP devices is based on exceptional fabrication precision of SNAP technology combined with the exceptionally low loss of silica and smoothness of the optical fiber surface.

Several fabrication methods of SNAP devices have been developed. They include $CO_2$ laser annealing [14-17], introduction of internal stresses by femtosecond laser inscription [18-20], variation of the fiber refractive index by local heating [21, 22], mechanical tuning by strong bending of optical fibers [23], exploring capillary fibers with a droplet inside [24], slow cooking of microresonators in the capillary fibers [25], as well as fiber tapering [26]. While most of these methods have exceptional precision, they allow the introduction of effective fiber radius variation (ERV) of the fiber by several nanometers only. In fact, the most flexible fabrication method using $CO_2$ laser exposure exploits compressive stresses frozen into optical fibers in the process of drawing. Release of these stresses allows to introduce the ERV which is limited by several nanometers [14-17]. Femtosecond laser inscription has a similar limitation [18-19]. Other less flexible methods allow much larger ERV but without precise control [20, 26].

For several applications it is necessary to develop a method which allows us to fabricate microresonators with a greater ERV ranging from nanoscale to microscale with high precision. For example, it has been shown that creation of broadband and low repetition rate frequency comb generators requires fabrication of SNAP bottle microresonator (BMR) having ERV of ~100 nm [27]. Increasing ERV allows us to fabricate resonance structures at the optical fiber surface having smaller axial dimensions. Examples of such structures include chains of coupled ring resonators theoretically investigated in [28, 29]. Overall, development of approaches which simultaneously enable introduction of nanoscale and microscale ERV of the optical fiber will significantly increase the flexibility of the SNAP technology and its possible applications.

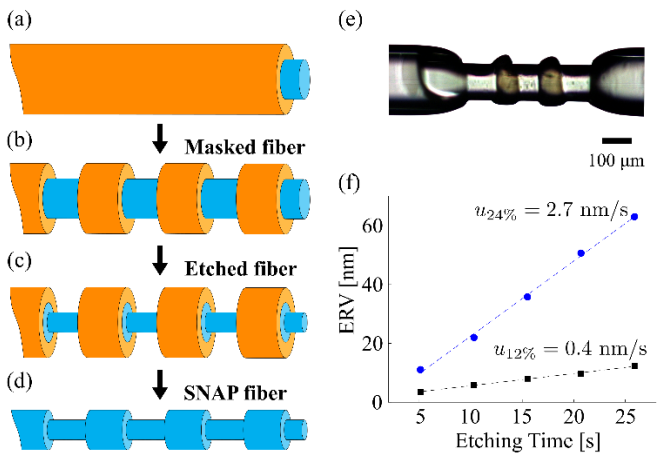

Fig. 1. Step-by-step lithographical fabrication approach. (a) uncoated fiber, (b) masked fiber fabricated with a focused $CO_2$ laser beam, (c) silica fiber etched in HF acid, and (d) fully uncoated silica fiber with series of SNAP BMRs introduced. (e) Optical microscope image of a 125 μm diameter fiber with partly removed coating. (f) ERV variation as a function of time for etching in 12% and 24% HF acid solutions.

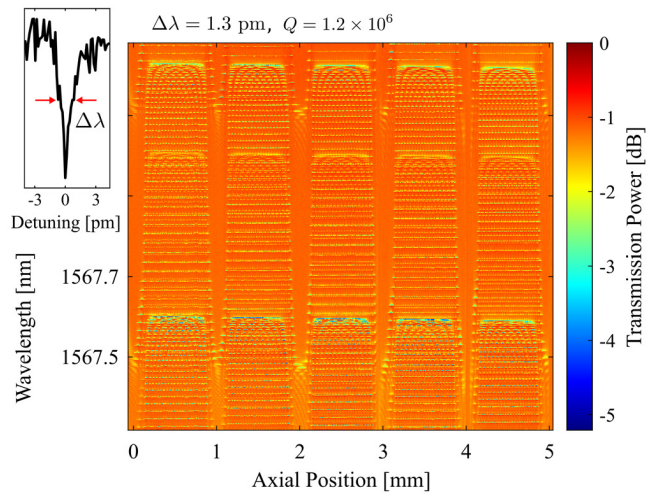

Fig. 2. The transmission power spectrogram of a chain of 5 SNAP BMRs fabricated at the surface of a 125 μm diameter optical fiber. Top left: the resonance spectrum used to determine the microresonator Q-factor.

Here, we present an approach which can solve this problem. The idea of our method is illustrated in Fig. 1 and follows, in part, the approach previously suggested in Ref [30], which, to our knowledge, has not been demonstrated experimentally. We developed a simplified lithographic process where the required ERV is introduced by wet chemical etching of an optical fiber using its coating removed from predetermined surface areas as a mask. In our proof of concept experiments we, first, used a focused $CO_2$ laser beam to remove fiber coating (Fig. 1(a) and (b)). The optical microscope image of a fiber with partly removed coating is shown in Fig. 1(e). In future development, a more accurate mask fabrication process can be implemented by exploiting methods of the fiber coating removal demonstrated previously [31-33]. In particular, excellent accuracy of the coating removal has been demonstrated with a vacuum ultraviolet laser in Ref. [33]. Next, the partly uncoated fiber was loaded into the hydrofluoric (HF) acid solution and etched for the duration of time needed to introduce the required ERV (Fig. 1(c)). To this end, we calibrated the etching process for different HF concentrations. In particular, Fig. 1(f) shows the dependence of ERV (measured as described below) on the etching time in 12% and 24% HF acid solutions, which corresponded to the etching speed of 0.4 nm/s and 2.7 nm/s. While in the applications described below we perform uniform etching of silica fibers, more general HF etching methods previously explored for fabrication of optical fiber tapers [34, 35] and microresonators [36-38] can be applied to fabricate BMRs with predetermined ERV. Next, the etched fiber sample was cleaned in deionized water and uncoated from the polymer mask in sulfuric acid solution (Fig. 1(d)). Finally, we created a SNAP fiber with a chain of high Q-factor rectangular BMRs by polishing the processed fiber sample with a travelling low-power $CO_2$ laser beam. In our experiments, we demonstrated the developed approach using 125 μm and 38 μm diameter silica fibers.

In our first experiment, we processed the standard 125 μm optical fiber as described above (Fig. 1) to fabricate a chain of 5 rectangular BMRs introduced with 1 mm period and 300 μm separation. These BMRs were characterized by their spectrogram which was measured using a transversely oriented tapered fiber with ~1 μm waist diameter (microfiber). The microfiber was scanned along the fiber under test periodically touching it at equally spaced positions as described previously (see, e.g. [14-17]). The spectrogram measured with Luna OVA shown in Fig. 2 is a 2D plot of transmission power as a function of wavelength and microfiber position along the axis of the fiber under test. The spectrogram shows a series of resonances corresponding to WGMs localized in these BMRs. The spectrogram bandwidth includes resonances corresponding to localized WGMs with different axial, azimuthal and radial quantum numbers. In particular, this spectrogram includes three cutoff wavelengths (thick yellow segments inside BMRs) with different azimuthal and radial quantum numbers of the fiber. Note that while the straight horizontal profile of cutoff wavelengths is clearly seen in the spectrogram of Fig. 2, the cutoff wavelength profile in between the BMRs is washed out and seen only partly. We partly attribute the latter effect to the imperfectness of the polymer mask edges noticeable in Fig. 1(e). We determined the spatial profile of BMRs by the rescaling relation between the ERV $\Delta r(z)$ along the fiber axis $z$ and cutoff wavelength variation $\Delta\lambda_c(z)$ in the form $\Delta r(z) = r_0(\Delta\lambda_c(z)/\lambda_c)$ where $r_0$ and $\lambda_c$ are the radius and cutoff wavelength of the fiber. Slight tilt of the BMR height visible in Fig. 2 is due to the natural variation of the ERV of the fiber [39]. Comparing the values of cutoff wavelengths in the areas within and outside of the BMRs, we estimate the introduced cutoff wavelength variation as 0.77 nm and ERV as 31 nm. The Q-factor of the fabricated microresonators, $Q = 1.2 \cdot 10^6$, (see inset in Fig. 2) was measured with the APEX optical spectrum analyzer (wavelength resolution 0.14 pm).

In our second experiment, we fabricated a 0.5 mm long BMR which is much deeper than those previously demonstrated in SNAP technology [14-25] and also than BMRs presented in Fig. 2. The spectrogram of this BMR is presented in Fig. 3. It was fabricated at the surface of a 38 μm diameter fiber and had the cutoff variation $\Delta\lambda_c = 24.2$ nm corresponding to 291 nm ERV, which is an order of magnitude greater than the ERV of SNAP microresonators fabricated by releasing stresses with a $CO_2$ laser or their introduction with a femtosecond laser [14-19]. The slope of the cutoff wavelength variations in Fig. 3 at the BMR edges, which determines the available contrast of our method, is 0.2 nm/μm. This is an order of magnitude greater than that achieved previously in [14-19, 21-23, 25-26] but less than the contrast achieved by the femtosecond laser cutting method [20] or droplet-induced BMRs [24]. In fabrication of this BMR, we had in mind the application of the developed method to the fabrication of broadband and low repetition rate frequency comb generators [27]. The design proposed in Ref. [27] required that the bandwidth of the axial modes with fixed azimuthal and radial quantum numbers is equal or greater than the free spectral range along the azimuthal quantum number. For the silica fiber with radius $r_0 = 19$ nm and refractive index $n = 1.44$ considered, at wavelength $\lambda = 1570$ nm, this free spectral range is $\lambda^2/(2\pi n r_0) \cong 14.3$ nm, i.e., smaller than $\Delta\lambda_c$. For the application of comb generation, the technique described should be modified to introduce a BMR with a parabolic shape. This can be done by application and extension of nonuniform HF etching methods [34, 35].

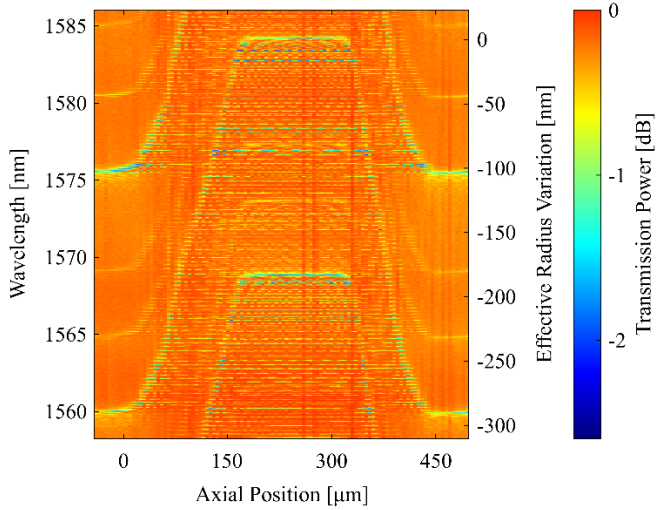

**Fig.** 3. The transmission power spectrogram of a 0.5 mm long BMR fabricated at the 38 μm diameter silica fiber.

In our third experiment, we used a 38 μm diameter fiber to fabricate a 5 mm long and 25 nm high rectangular BMR microresonator which spectrogram is shown in Fig. 4(b). The goal of this experiment was to investigate the performance of this resonator as a miniature delay line. In particular, we demonstrated that this BMR can be used as a *dispersionless* delay line provided that the pulse center wavelength is separated from the cutoff wavelengths of the fiber.

The theoretical baseline of our third experiment was as follows. It is known that the smallest and slowest dispersionless delay line can be realized using a BMR with a semi-parabolic shape [16], while deviation from the parabolic shape introduces dispersion. It is also known that the slowest axial propagation of WGMs is achieved near the cutoff wavelengths of optical fibers and BMRs (see e.g., [20]). The behavior of the WGM propagation constant near the cutoff wavelength $\lambda_c(z)$ of a fiber is determined as [14]

$$\beta(\lambda, z) = \frac{2^{3/2}\pi n}{\lambda_c^{3/2}}(\lambda_c(z) - \lambda)^{1/2} \quad (1)$$

which yields for the WGM axial group velocity $v_g$ and group velocity dispersion (GVD):

$$v_g = -\frac{2\pi c}{\lambda_c^2}\left(\frac{\partial\beta}{\partial\lambda}\right)^{-1} = 2^{1/2}\frac{c}{n}\left(\frac{\lambda_c(z)-\lambda}{\lambda_c}\right)^{1/2}, \quad (2)$$

$$GVD = \left(\frac{\lambda_c^2}{2\pi c}\right)^2 \frac{\partial^2\beta}{\partial\lambda^2} = \frac{-n\lambda^{5/2}}{2^{5/2}\pi c^2(\lambda_c(z)-\lambda)^{3/2}}. \quad (3)$$

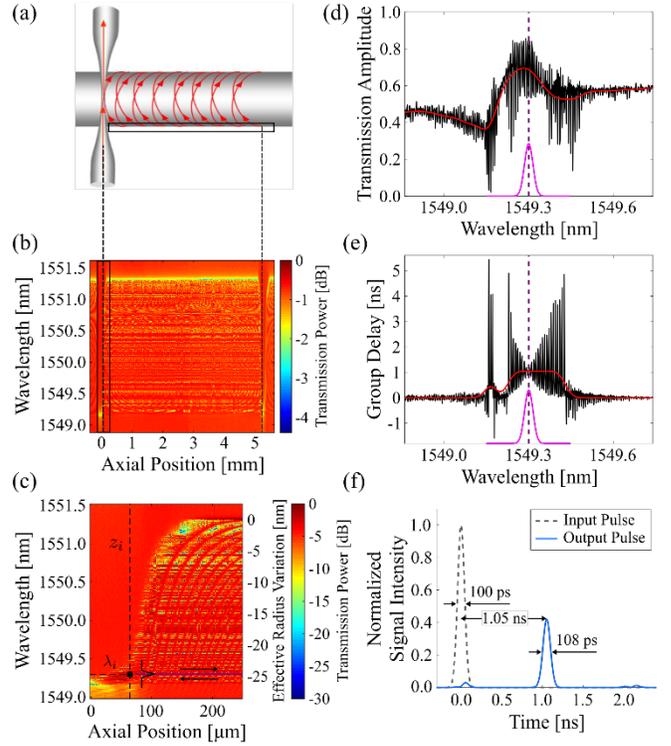

Fig. 4. (a) Illustration of the input-output microfiber position with respect to a SNAP microresonator delay line. (b) The spectrogram of a 5.3 mm long BMR fabricated at a 38 μm fiber. (c) A section of spectrogram (b) magnified near the position of the input/output microfiber. (d), (e) Transmission amplitude and group delay spectra (black lines) and averaged spectra (red lines) at the microfiber position corresponding to impedance matching at $\lambda_i = 1549.3$ nm with co-located 100 ps pulse bandwidth depicted (magenta line). (f) Propagation of the 100 ps pulse with central wavelength $\lambda_i$.

From Eq. (3), the GVD is large for wavelengths in close proximity of the cutoff wavelength $\lambda_c(z)$ and vanishes away from $\lambda_c(z)$. For this reason, pulses with central wavelength $\lambda_p$ close to $\lambda_c(z)$ experience large dispersion. However, pulses having the central wavelength $\lambda_p$ away from $\lambda_c(z)$ and width $\Delta\lambda_p \ll |\lambda_c(z) - \lambda_p|$ will experience small dispersion in the process of propagation. Thus, in this experiment we first determined the wavelength $\lambda_i = 1549.3$ nm at the bottom of our BMR and microfiber position $z_i$ corresponding to the impedance matching condition. Under this condition, light entering the BMR exits it completely after a single round trip [16]. The magnified section of the transmission power spectrogram near $z_i$ is shown in Fig. 4(c). The transmission amplitude and group delay spectra at position $z_i$ are shown in Figs. 4(d) and (e). The group delay spectrum has vanishing oscillations at the impedance matching wavelength $\lambda_i$ as expected. From Fig. 4(e), we find the average group delay in the vicinity of $\lambda_i$ equal to 1.05 ns. Theoretically, the group delay for our BMR is determined as its double length $2L = 10.6$ mm (due to the fact that the pulse entering the resonator makes a full roundtrip before exiting) divided by the pulse velocity $v_g$ calculated from Eq. (2) for $\lambda_c = 1551.3$ nm (Fig. 4(c)), $\lambda = \lambda_i = 1549.3$ nm and $n = 1.44$. From here, we find the pulse velocity $v_g = 1.06 \cdot 10^7 m/s \cong 0.035c$ and the delay time $\tau = 2L/v_g = 1.002$ ns in good agreement with the experimental data. Finally, Fig. 4(f) shows the propagation of a Gaussian pulse with FWHM $\Delta T_0 = 100$ ps and central wavelength coinciding with the impedance matching wavelength $\lambda_p = \lambda_i = 1549.3$ nm, which was calculated using the experimental data shown in Figs. 4(d) and (e). Good agreement is observed with local average values for transmission power and group delay (red lines in Fig. 4(d) and (e)). For an ideally rectangular BMR, the FWHM of a Gaussian pulse after propagating the roundtrip distance $2L$ is $\Delta T = \Delta T_0 (1 + 4L^2 \cdot GVD^2/\Delta T_0^4)^{1/2}$ [40]. Calculating GVD with Eq. (3) for our parameters, we find $\Delta T = 1.001 \Delta T_0$ i.e., negligible dispersion. From Fig. 4(f) we find a small but noticeably greater pulse broadening, $\Delta T = 1.08 \Delta T_0$, which we primarily attribute to the finite slope of the BMR at the edges (Fig. 4(c)).

In conclusion, we have demonstrated a method for fabrication of BMRs at the surface of an optical fiber using its selectively removed coating as a mask for HF chemical etching. The developed technique allowed us to fabricate individual and multiple BMR structures with arbitrarily large ERV, which was not possible previously in SNAP technology. In one of the examples considered, we investigated a miniature rectangular 5.3 mm BMR delay line, where, to avoid dispersion, the central wavelength of propagating pulse was chosen separated from the cutoff wavelengths. We experimentally showed that, for sufficiently large separation and relatively small pulse bandwidth (100 ps in this example), a pulse, which was delayed by 1 ns, experienced negligible dispersion. In addition to vanishing dispersion, the critical advantage of exploring BMRs away from the fiber cutoff wavelengths as delay lines consists in the immunity of their performance to the shape of a BMR provided that it is smooth enough and does not cause reflections. These advantages are achieved at the expense of increasing the group velocity and, thus, the size of the delay line for the same delay time. Overall, we believe that the presented proof of concept demonstrations can be significantly improved and extended by using more precise methods of fiber coating removal and nonuniform etching as well as by combination of the developed approach with other fabrication methods of SNAP technology.

**Funding.** Engineering and Physical Sciences Research Council, (EP/P006183/1), Horizon H2020-MSCA-COFUND (713694), Wolfson Foundation (22069).

**Disclosures**. The authors declare no conflicts of interest.